\documentclass{article}
\usepackage{graphicx} 
\usepackage{comment}
\usepackage{amsmath}
\usepackage{textgreek}
\usepackage{geometry}
\usepackage{float}
\usepackage{authblk}
\usepackage{lineno}
\title{Interferometric modal splitting enables a broadband, dual-polarization on-chip spectrometer}
\author[1]{Karl Johnson}
\author[1]{Vladimir Fedorov}
\author[1]{Dmitrii Belogolovskii}
\author[1]{Andrew Grieco}
\author[1]{Noah A. Rubin}
\author[1]{Yeshaiahu Fainman}
\affil[1]{Department of Electrical and Computer Engineering, University of California San Diego, La Jolla, CA, USA}
\date{February 2025}
\usepackage[backend=biber,style=nature]{biblatex}
\usepackage{hyperref} 

\begin{document}

\maketitle

\section{Abstract}

The modal dispersion of waveguides often limits integrated photonic devices to operation with a single polarization state. This presents a challenge for sensing and spectroscopy applications, which often require polarization diversity over wide bandwidths with high throughput. Here, we show that an unmodified thermally-driven silicon photonic Fourier transform spectrometer exhibits a polarization-separating effect in the frequency domain, even though only one polarization-insensitive detector is used. Using this effect, we experimentally demonstrate a simple on-chip spectrometer capable of extracting two-polarization spectra over a wide 1480-1630 nm bandwidth with a greater than 20 dB polarization extinction ratio. These specifications would be highly challenging to achieve using existing, conventional on-chip polarization-splitting techniques. We additionally demonstrate several improvements in calibration and testing that improve the performance of on-chip Fourier transform spectrometers even in the single-polarization case. The ``interferometric mode splitting" principle which this spectrometer exemplifies is general to various on-chip spectrometer architectures, other spatial modes, and technologies other than thermally-driven Fourier transform spectrometers. Interferometric mode splitting shows promise as a general approach for robust and fundamentally broadband detection of orthogonal modes in guided-wave sensing.

\section{Introduction}
    Developments in integrated photonics have revolutionized optical communications, and have brought new capabilities for basic research \cite{siew2021siliconphotonicsreview,miri2019exceptional,yu2019boundstates}. The unique physics and size advantages of on-chip optics have also stimulated substantial interest in optical sensing applications, both in miniaturizing existing technologies and in realizing new ones \cite{rank2021oct,zhang2022lidar,lai2020gyroscope, passaro2012ringresonatorsensors}. In recent years, miniaturized on-chip optical spectrometers have seen notable advancements \cite{yang2021sciencereview}. The resolution, bandwidth, and throughput of on-chip spectrometers have dramatically increased, while the footprint of such devices has continued to shrink \cite{xu2023photonicmolecule,yao2023programmable,yao2023reconfigurable}.
    
    However, transitioning optical systems to a chip often introduces unique challenges. Most integrated photonic devices are designed to operate using only one waveguide mode, corresponding to a single spatial mode with one polarization. In optical communications, mature solutions exist for supporting two polarization modes, through a combination of polarization-splitting components \cite{fukuda2008diversity,taillaert2003gratingcoupler} and signal processing to account for polarization fluctuations in incoming light and in the fiber-optic channel \cite{dai2013polarizationmanagement,shahwar2024polarizationmanagement}. 
    
    Meanwhile, for optical sensing applications - such as optical spectroscopy - in which light may move on and off the chip and interact with the environment, this single-mode limitation is a more serious issue. Practical applications of optical spectroscopy demand the highest bandwidth and throughput possible. Supporting only one waveguide mode yields a substantial throughput limit for coupling extended light sources on-chip \cite{hudson1974modecoupling}. Additionally, single-polarization operation may lead to ambiguities or artifacts in the measurement if incoming light has a spectrally-varying polarization state. One method to achieve multi-mode operation is to couple off-chip light into several waveguide modes and use an on-chip mode or polarization splitter to route the light to multiple copies of the on-chip device \cite{fukuda2008diversity}. However, this increases system complexity, and designing such on-chip mode splitters to have a wide bandwidth is highly challenging. For example, on-chip broadband polarization splitters typically exhibit poor polarization extinction ratios (PER's) of just 12-16 dB over the wide bandwidths of interest for spectroscopy applications \cite{yin2017broadbandsplitter,wang2014broadbandsplitter,zhao2018broadbandsplitter}.

    Guided wave technologies such as integrated photonics struggle to simultaneously operate with more than one mode for a simple reason: every mode in a waveguide has different properties, all of which are highly wavelength-dependent. Substantial design effort can yield devices that behave similarly for multiple modes at specific wavelengths, but such efforts to ``correct" the differences between modes are inherently narrowband \cite{dai2013polarizationmanagement}. On the other hand, in the field of free-space polarization optics, highly broadband polarization splitters have been constructed for hundreds of years \cite{goldstein2017polarizedlightbook}. This bandwidth is derived by harnessing the differences in absorption or refractive index between polarizations that remain valid over broad bandwidths. This suggests that harnessing the differences between modes rather than laboring to ``correct'' them is a more robust approach for broadband management of several modes in integrated photonics. Indeed, existing integrated mode splitters leverage the difference in effective index between modes for their operation; however, mode splitters are coupled-mode devices relying on phase-matching, which is also a fundamentally bandwidth-limited technique \cite{yariv1973coupled}.

\begin{figure}
    \centering
    \includegraphics[width=100mm]{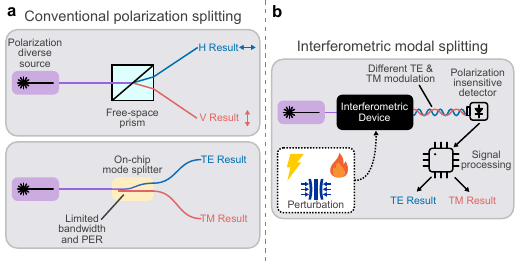}
    \caption{a) Typical methods of polarization or mode diversity in optical systems require beam splitters, which suffer from poor bandwidth and PER on-chip. b) In the method proposed here, modes share a common physical channel in which they experience different optical responses to a common perturbation (e.g., electric field, stress, or heat). These different responses enable mode separation in post-processing, avoiding the challenges inherent to physical separation over broad bandwidths.}
    \label{fig:enter-label}
\end{figure}

    In contrast to the narrowband nature of coupled-mode devices, interferometric structures can convert phase variations between modes into intensity variations (fringes) across very broad bandwidths. In this work, we propose that interferometric photonic sensing architectures can utilize this inherent advantage to naturally separate modes over broad bandwidths (Fig. 1). To demonstrate this general principle of ``interferometric modal splitting'' (IMS), we show that temperature-derivative modal dispersion (TDMD) causes frequency-domain polarization splitting in a thermally actuated Fourier transform spectrometer. This phenomenon manifests in the waveguides themselves, and simple signal processing is used to extract the two spectra from data measured using a polarization-insensitive detector. There are no changes to the spectrometer design or architecture from previous works \cite{souza2018fourier, li2021widewaveguides}. The demonstrated spectrometer is capable of simultaneously measuring spectra of two orthogonal polarizations with a wide bandwidth (1480-1630 nm), high extinction ratio (approximately 20 dB), and moderate resolution ($R=\frac{\lambda}{\Delta \lambda}\approx 600$ for the TE mode and $R\approx 1500$ for the TM mode, or $\Delta\lambda =$ 2.5 and 1 nm, respectively). The measured bandwidth and extinction ratio reach the limits of our test equipment. Beyond the demonstration here, IMS easily generalizes to orthogonal modes other than polarization, more advanced spectrometer architectures, and modal differences other than TDMD. 

\section{Results}

\subsection{Temperature-derivative modal dispersion and frequency-domain mode splitting}
    In an interferometer with a monochromatic input of frequency $\nu_0$, the fluctuations in output intensity (interferogram) are sinusoidal with the optical path length (OPL) difference between the two sides (Fig. 2a-b),
    \begin{equation}
        I_\text{out} = I_0 \cos(\phi_2 - \phi_1) = I_0 \cos(k_0 (\text{OPL}_2 - \text{OPL}_1)) = I_0 \cos(k_0 \Delta \text{OPL}),
    \end{equation}
    where $I_0$ is the intensity of the monochromatic input, $k_0$ is the free space wavenumber $\frac{2\pi \nu_0}{c_0}$, and $\text{OPL}_{1,2}$ are defined in Fig. 2a. Note that here we drop the constant intensity offset of this interferogram (which is required for the intensity to remain positive) for brevity. In this work, we consider interferometers in which the path length $\Delta \text{OPL}$ is varied by use of heat and the thermo-optic effect. In the case of a thermally-driven, balanced interferometer with $L_1 = L_2$ and a linear thermo-optic response, the measured interferogram is dependent only on $\Delta T = T_2 - T_1$, with $\Delta\text{OPL} = L \frac{dn_\text{eff}}{dT}\Delta T$ such that
    \begin{equation}
        I_\text{out}(\Delta T) = I_0\cos(\frac{2\pi \nu_0}{c_0} L \frac{dn_\text{eff}}{dT}\Delta T).
    \end{equation}
    This has a frequency with respect to $\Delta T$ of
    \begin{equation}\label{eq:temp_freq}
         f_T = \frac{L}{c_0}\nu_0 \frac{dn_\text{eff}}{dT}\Big|_{\nu = \nu_0}.
    \end{equation}
    The one-to-one relationship between frequency in the interferogram and optical frequency is the classic operating principle of a Fourier transform spectrometer. As interferograms of each wavelength component add linearly, taking the Fourier transform of the interferogram yields a scaled version of the optical spectrum \cite{davis2001ftsbook} (Fig 2c-e). In the two-polarization case, however, each polarization mode has a different modal overlap with the waveguide core and cladding materials and therefore different $\frac{dn_\text{eff}}{dT}$ (Fig. 2j). If a strictly monochromatic source with some power in both polarization components is incident on the chip, a polarization-insensitive detector will measure a linear sum of the two interferograms (Fig. 2g). Taking the Fourier transform with respect to the temperature difference will yield two defined peaks (Fig. 2h). Each corresponds to one of the two polarizations: the polarizations have ``split'' in the frequency domain, even though both were measured simultaneously on the polarization-insensitive detector. In other words, the polarization mode dispersion of the waveguide itself encodes polarization into the interferogram period in a way that can be later separated.

        \begin{figure}[H]
            \centering
            \includegraphics[width=88mm]{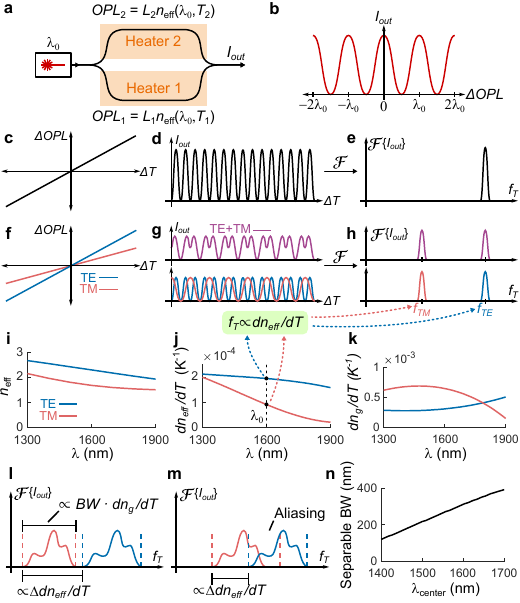}
            \caption{Polarization-separating FTS concept for narrowband and broadband inputs. (a) Thermally-actuated Mach-Zhender interferometer (MZI) architecture for the FTS, with a narrowline laser input as an example; OPL = optical path length. (b) The output fringes of an interferometer are sinusoidal when plotted versus $\Delta$OPL = OPL$_2$ - OPL$_1$. (c) In the single wavelength case, a linearly varying $\Delta$OPL yields (d) a sinusoidally varying output intensity with respect to the temperature difference between the interferometer arms, (e) the Fourier transform of which is a single peak: a reconstruction of the input laser spectrum. (f) For a two-polarization FTS, the different $\Delta$OPL experienced by each polarization due to TDMD yields (g) two superimposed sinusoids that are distinct peaks in (h) the Fourier domain. (i) Effective indices, (j) effective index temperature derivatives, and (k) group index temperature derivatives of the TE and TM modes in a 450 x 220 nm waveguide. The temperature domain frequencies of the TE and TM fringes for a narrowband input $\lambda_0$ ($f_{TM}$ and $f_{TE}$) shown in (h) are directly proportional to the temperature derivative (j) at $\lambda_0$.
            (l) For strong TDMD, broadband spectra can separate without overlap in the frequency domain; BW = optical bandwidth. (m) In the case of weak TDMD, the two polarizations may overlap in the frequency domain, causing aliasing. (n) Maximum spectral bandwidth over which both polarizations can be separated in a 450 x 220 nm waveguide without aliasing. The separable spectral bandwidth is a function of the center wavelength of the spectral bandwidth.}
        \end{figure}

    However, note that Eq. \ref{eq:temp_freq} indicates the splitting of modes does \emph{not} arise from the conventional modal dispersion of $n_\text{eff}$ (Fig. 2i), but rather what we dub `temperature-derivative modal dispersion' (TDMD), i.e., differences in $\frac{dn_\text{eff}}{dT}$ between modes (Fig. 2j). Derivative modal dispersions such as TDMD have previously been shown to be important in evanescent index sensors \cite{johnson2022ringtoc, liu2016ring2pol, ding2019ring2pol} and also have utility in lab techniques involving polarization-maintaining fiber \cite{thorlabsPM}. Despite this, separating modes using derivative modal dispersion for sensing purposes, to the authors' knowledge, has not been previously discussed. 
    
    A striking property of the frequency domain polarization splitting in the case of a single narrowband input is wavelength diversity. Outside of discrete wavelengths where the curves in Fig. 2j cross, TDMD is nonzero at every wavelength which supports the modes, yielding splitting for any narrowline input. This is in contrast to coupled-mode polarization splitters, where a finite amount of power from each mode will always `leak' into the inappropriate output waveguide, with the extinction ratio approaching zero outside the design wavelength range.

\subsection{Separation of broadband sources}
    For non-monochromatic inputs, each polarization has a support of some finite width in the fringe frequency domain ($f_\text{T}$ domain), proportional to the bandwidth of the source in the optical frequency domain (Fig. 2l). Since the two polarizations are only separated in the fringe frequency domain by a finite amount, for wide optical bandwidths the fringe domain support corresponding to each polarization will overlap. In this case the polarizations cannot be unambiguously separated (Fig. 2m), yielding aliasing between the two polarizations.

    As discussed in the previous section, TDMD is responsible for the polarization splitting of a single wavelength in the fringe frequency domain. Intuitively, then, stronger TDMD generally allows for broader optical bandwidths to be reconstructed without aliasing. However, the wavelength dependence of the TDMD substantially impacts the optical bandwidth that can be separated without aliasing. In Supplementary Section \ref{supp:separability}, we derive a general condition to determine the  optical bandwidth over which polarizations will remain separable. In the simplified case of a waveguide with a constant (but not necessarily identical) $\frac{dn_\text{g}}{dT}$ for both polarizations (with $n_g$ being the usual group index), the separability condition reduces to 
        \begin{equation}\label{eq:bandwidth}
             \frac{\nu_2}{\nu_1} < 1 + \frac{\frac{dn_\text{eff}}{dT}\big|_{TE,\nu_1} - \frac{dn_\text{eff}}{dT}\big|_{TM,\nu_1}}{\frac{dn_\text{g}}{dT}\big|_{TM}},
        \end{equation}
    where the separable optical frequency bandwidth is $[\nu_1, \nu_2]$. The term in the numerator on the right-hand-side corresponds to the TDMD as previously described. This term is proportional to the distance between the lower-frequency edge of the two polarizations' support in the frequency domain, so larger values increase the separable bandwidth (Fig. 2l). Meanwhile, the group index temperature derivative $\frac{dn_\text{g}}{dT}$ (which is related to the wavelength derivative of the TDMD) determines the width of each polarizations' support in the frequency domain, where a higher value 'spreads out' the spectrum. As such, a higher $\frac{dn_\text{g}}{dT}$ of the leftmost polarization actually reduces the separable bandwidth, so it appears in the denominator of Equation \ref{eq:bandwidth}. Notably, the separable bandwidth in Eq. \ref{eq:bandwidth} is fractional and wavelength-dependent - as such, the choice of one bound of the operational bandwidth shifts the other bound of the operational bandwidth as well. Figs. 2j,k,n show the wavelength-dependent $\frac{dn_\text{eff}}{dT}$, $\frac{dn_\text{g}}{dT}$, and separable bandwidth for the 450 $\times$ 220 nm silicon-on-insulator (SOI) waveguide profile used in this work (calculated using the more accurate general expression given in Supplementary Section \ref{supp:separability}). At long wavelengths, the TDMD shown in Fig. 2j is strong, yielding a predicted separable bandwidth approaching 400 nm. 

   The group index temperature derivative $\frac{dn_\text{g}}{dT}$ also directly impacts spectrometer resolution. For a spectrometer with delay lines of length $L$ and a maximum temperature difference between arms $\Delta T_\text{max}$, the resolution at a given wavelength is of the form
        \begin{equation}\label{eq:resolution}
            \Delta \nu |_{\nu_0} \propto \frac{c_0}{2L\Delta T_\text{max} \frac{dn_\text{g}}{dT}|_{\nu_0}}.
        \end{equation}
    This expression has been presented previously and is very similar to the well-known expression for the resolution of a Fourier transform spectrometer (in particular, in that it is inversely proportional to the total differential OPL) \cite{kita2018fts, davis2001ftsbook}. We perform a more general derivation of the resolution in Supplementary Section \ref{supp:ng_resolution}. As shown in Fig. 2k, the two modes in a standard SOI waveguide have a substantially different group index derivative, and as such the two polarizations will have a substantially different resolution in the Fourier transform spectrometer.

    In practice, there are nonlinearities both in the temperature-domain tuning and wavelength-domain dispersion of the indices, which not only complicate the previous expressions, but must be dealt with to obtain a clean spectral reconstruction \cite{souza2018fourier}. In this work, we employ push-pull tuning of the interferometer, in which the temperature of both interferometer arms are changed simultaneously in opposite directions. This is a novel improvement of on-chip Fourier transform spectrometers that removes the strongest nonlinearities of the thermal tuning and substantially improves the performance of our spectrometer (Supplementary Section \ref{supp:push_pull}). Corrections for the remaining dispersion and thermo-optic nonlinearity terms are performed using a nonuniform discrete Fourier transform (NUDFT) technique which performs nonlinear temperature-domain scaling, the Fourier transform, and frequency-domain scaling. More details regarding the calibration of the spectrometer and the NUDFT extraction technique can be found in Supplemental Sections \ref{supp:nufft} and \ref{supp:data_processing}.

\subsection{Experimental results}
    To experimentally demonstrate the proposed dual-polarization spectrometer, we fabricated several spectrometers on a standard 220 nm silicon-on-insulator waveguide platform (Fig. 3a, see Methods for further information). In Supplemental Section \ref{supp:widths}, we show results from several spectrometers demonstrating that the resolution and separable bandwidth of the spectrometer can be varied substantially by adjusting the waveguide width. For the demonstrations here, we show results from a spectrometer with a 450 nm waveguide width, which we observed to have the best trade-off between resolution, separable bandwidth, and nonidealities arising from sidewall roughness.

    Figs. 3b-c show a representative example of polarization splitting, and are the experimental equivalent to Fig. 2g-h. A 1540 nm laser source is coupled to the chip at a 45-degree polarization angle (with respect to the principal axes of the waveguide). The output power is measured as a function of time as the power through heaters on the two interferometer arms are swept linearly (push-pull, Fig. 3b). Taking a fast fourier transform (FFT) yields two clear spikes in the frequency domain, one for each polarization component of the incoming light. 

    Figs. 3d-e show FFT's of interferograms collected when a polarization-aligned laser is input at wavelengths spanning 1480-1630nm (the maximum we can achieve with our equipment). The broadband polarization separation can be seen clearly, with the two polarizations occupying two non-overlapping ranges in the frequency domain. Also visible is the different spacing between peaks for TE and TM modes due to the different group index temperature coefficient of the two modes. Also of note here is the broadening of the peaks in the frequency domain due to slight nonlinearities in the thermo-optic tuning, which are shown in more detail in Fig. 3f.  When this nonlinearity is corrected for using the aforementioned NUDFT technique (Supplemental Section \ref{supp:nufft}), the broad peaks become narrow transform-limited spikes in the frequency domain (Figs. 3g-h).

    The noise floor visible in the nonlinearity-corrected FFT in Figs. 3g-h suggest that the spectrometer has a dynamic range of approximately 20-25 dB. Notably, no crosstalk is visible at all between polarizations, as no peaks above the noise floor occur in the TE FFTs at TM frequencies, or vice versa. This yields an initial conservative estimate of the polarization extinction ratio (PER) at 20-25 dB (limited by the spectrometer dynamic range), which is discussed further later in this work. There is one artifact visible in 3h, namely additional peaks at higher frequencies approximately -20 dB weaker than the main peaks. These are at twice the frequency of the main peaks, and result from an unidentified nonlinearity in the detector or elsewhere in the system.
        \begin{figure}[H]
            \centering
            \includegraphics[width=180mm]{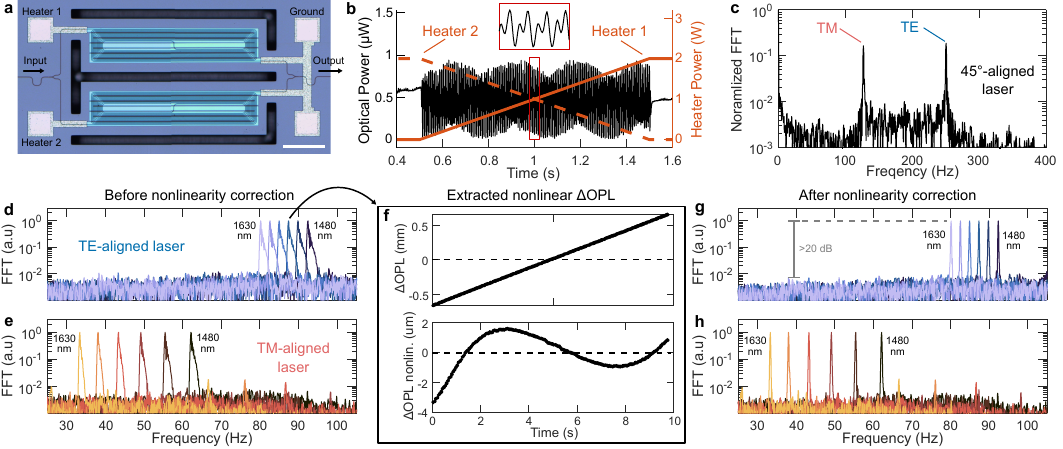}
            \caption{Characterization of the fabricated device. a) Optical microscope image of the fabricated spectrometer, scale bar length is 250 um. b) Interferogram measured of a 1540 nm laser source incident on the spectrometer with a 45 degree polarization, which couples into both polarization modes. The power delivered to each heater versus time is also plotted for reference to show the push-pull control. c) FFT of the interferogram in b, demonstrating two clear peaks associated with the two polarization modes. Note that the heaters were only driven to a maximum power of 2 W in b-c to minimize nonlinearities for illustrative purposes. All other results shown in this paper drive the heaters to 6 W, which triples the spectrometer resolution, but introduces unavoidable nonlinearities in the thermal tuning. d) FFT's of interferograms collected for a polarization-aligned laser set to various wavelengths spanning the 1480-1630 nm wavelength range for the TE mode and e) TM mode. f) OPL difference in the interferometer over the duration of the sweep and its deviation from a linear response. Though the optical path length difference appears very linear when viewed with an absolute scale (top pane), the nonlinearity is more obvious when the deviation from a linear response is plotted (lower pane). The method used to extract this nonlinear response is discussed in Supplemental Section \ref{supp:data_processing}, as characterizing this nonlinearity is critical to calibrating the spectrometer. g) FFT's of interferograms for the TE and h) TM modes performed when the nonlinearity shown in f) is corrected for.}
        \end{figure}

    Figs. 4a-h demonstrate the spectrometer's capability to reconstruct spectra with diverse polarization and spectral characteristics. The spectrum extraction and amplitude normalization method used in these plots is covered in Supplementary Section \ref{supp:data_processing}. In all these examples, the two polarization spectra shown are incident on the spectrometer simultaneously, and are reconstructed from a single interferogram measurement. As illustrated in the experimental schematic Fig. 4i, we perform reference measurements using a benchtop optical spectrum analyzer (OSA) and free-space polarizer with a high PER (see Methods). Spectra measured include cross-polarized pairs of laser sources (a,e), polarized and unpolarized broadband sources (b,c), and combinations of broadband/narrowline sources of various polarization states (f,g). The sources measured in Figs. 4d and 4h have a spectrally-varying polarization state generated by passing a linearly polarized broadband source through a short polarization-maintaining fiber aligned at a 45 degree angle (Supplemental Section \ref{supp:setup}). As the polarization-maintaining fiber has slightly different effective indices for its two principal modes, it acts akin to a thick, multi-order waveplate, yielding spectral polarization fringes in the light exiting the fiber.

    Figs. 4j-k demonstrate the spectrometer resolution on pairs of closely-spaced lasers. This plot shows clearly that the TE mode has a worse resolution (approx 2.5 nm) than the TM mode (1 nm) because the group index temperature coefficients are different (Supplemental Section \ref{supp:ng_resolution}). The spectrometer also faithfully reconstructs the relative amplitudes of the two lasers, which are not equal due to the imperfect splitting ratio of the fiber coupler used to combine lasers. The low signal in cross-polarization reconstructions here also indicates the high PER of the spectrometer. In Supplemental Section \ref{supp:er}, we further discuss the polarization extinction ratio in these measurements as well as that of Fig. 4b. A dynamic-range-limited PER of 16-20 dB is seen in measurements of TE-aligned sources, and TM-aligned sources exhibit a PER of over 20 dB, limited only by our equipment. Finally, the results in Fig. 4l show no proximity effects when two cross-polarized lasers are scanned over each other, serving as another demonstration that the polarization reconstructions are completely independent. 
            
    \begin{figure}[H]
        \centering
        \includegraphics[width=180mm]{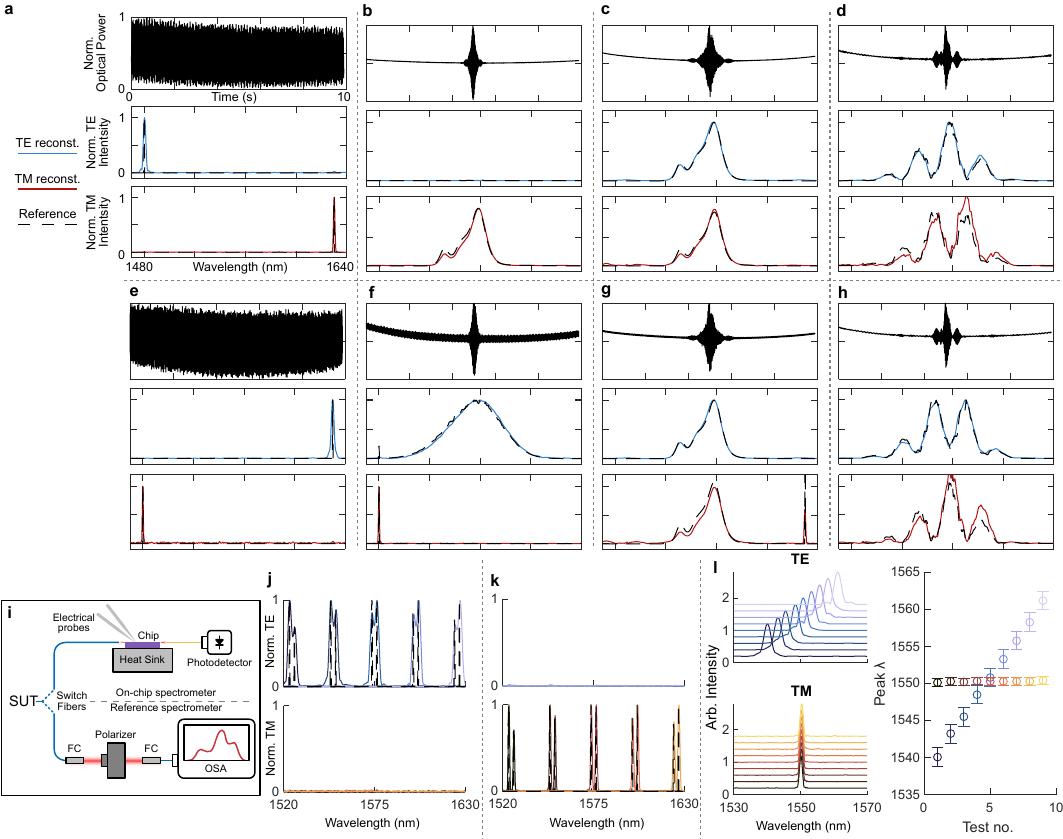}
        \caption{Broadband and polarization diverse spectral reconstructions. a-h) Demonstration of the spectrometer on a variety of two-polarization spectra. The top pane in each subfigure shows the measured interferogram, while the two panes below show the TE and TM reconstructions (colored lines) overlaid with their corresponding reference measurements (dashed black line). The axes and scaling are the same on all plots, and are removed from b-h) for clarity. Details on the experimental configurations to generate the spectra in each plot are given in Supplementary Section \ref{supp:setup}. i) Experimental setup used to obtain spectral measurements from the FTS as well as reference measurements from a benchtop spectrometer. SUT, spectrum under test; FC, fiber collimator, OSA, optical spectrum analyzer. j) Reconstructions of 2 closely-spaced lasers both aligned to the TE or k) TM modes across a wide bandwidth. Each color is a different experiment. The separation between lasers is kept at 3 nm as the two lasers are tuned to 5 positions across the bandwidth shown. The narrower 1520-1630 nm bandwidth in this plot is due to limits in the wavelength range of the second tunable laser. l) Numerous overlaid extractions for two cross-polarized lasers as the TE laser is swept while the TM laser remains fixed. Each color is a separate experiment; the right pane plots the center wavelength for each experiment (error bars correspond to the full-width-half-max of the reconstructed spectra).}
    \end{figure}
\section{Discussion}

     The two-polarization capabilities of the spectrometer presented here have several attractive features. Our experimental demonstration of the spectrometer bandwidth is limited by the tunable laser used for testing. In Supplementary Section \ref{supp:widths}, we show that our experimental results agree well with simulation, which predicts a wide 270 nm bandwidth for operation centered around 1550 nm. However, the demonstrated spectrometer also has some drawbacks. For one, it has a large footprint (800 x 1750 um), high power consumption (6 W), and only moderate resolution (2.5 nm on the TE mode). State-of-the-art computational on-chip spectrometers with substantially superior single-mode performance have been previously demonstrated \cite{yao2023programmable,yao2023reconfigurable,xu2023photonicmolecule}, though these spectrometers also have drawbacks in calibration complexity and robustness of spectral reconstruction.  Another drawback of the spectrometer here for practical use is the differing spectral resolution across the two polarizations. In an ideal dual-polarization spectrometer, the two polarization spectra can be combined to yield more general polarization information about the spectrum (e.g. added for total polarization-insensitive power). Spectral features resolved differently by the two polarization channels (such as those in Figs. 4j-k) will distort such combined measurements, and as such the higher-resolution spectrum must be blurred or downsampled before combined analysis can be performed. Finally, the dynamic range of the demonstrated spectrometer is quite limited in comparison to standard benchtop spectrometers, and is substantially lower than the dynamic range of the detector (at least 40 dB). We believe the dynamic range limitation is due to parasitic reflections from sidewall roughness in the 3-cm-long interferometer arms (Supplemental Section \ref{supp:er}).

    Fortunately, this spectrometer serves as only one example of the general principle of IMS introduced in this work. IMS is applicable to most on-chip spectrometer architectures, which is discussed in more detail in Supplemental Section \ref{supp:other_spectrometers}. Critically, this discussion indicates that two-polarization operation could likely be realized in numerous previously-demonstrated spectrometers with few or even zero hardware modifications. As computational spectrometers are known to be capable of high resolutions and high dynamic range, we believe that other researchers will be able to rapidly demonstrate dual-polarization computational spectrometers with performance far surpassing that demonstrated in this work by utilizing samples already demonstrated in previous publications.
    
    Dual-polarization spectrometers such as the one presented here have practical importance for a variety of spectrometry applications. Many spectrometry applications do not require polarization information; in these situations, a two-polarization spectrometer is still useful as summing the two polarization channels can yield polarization-insensitive results free from undesired spectropolarimetric effects (e.g., polarization fringes like those in 4d-4h). Other applications do benefit from dual-polarization information, such as measurements of Raman depolarization ratios and circular dichroism \cite{allemand1970depolarization, miles2021circdichroism}. For other applications, dual-polarization information is not sufficient and better understanding of the polarization state is required. Many of these applications, such as remote sensing of atmospheric clouds and aerosols \cite{dubovik2019aerosols}, require only degree-of-linear-polarization information, which can be performed using channeled-spectrum spectropolarimetry \cite{oka1999channeledspectrum, snik2009channeledspectrum}. This technique uses only a series of waveplates and a dual-polarization spectrometer, and as such is a good fit for on-chip spectrometers using interferometric modal splitting. Finally, for full-Stokes spectropolarimetry, 2- or 3-shot measurements would be required if a dual-polarization detector as demonstrated here were used along with additional, reconfigurable waveplates \cite{chipman2018polarizedlightbook, goldstein2017polarizedlightbook}. To adapt the two-polarization spectrometer demonstrated here for full-Stokes spectropolarimetry, waveguide technologies with an asymmetric modal profile could be used to perform the requisite polarization transformations on-chip required to fully sample the Poincare sphere \cite{dai2018asymmetricwaveguides, lin2019fullstokes}.

    While interferometric modal splitting is only experimentally demonstrated here using thermal tuning and polarization modes, it is nonetheless capabable of separating any orthogonal modes that have sufficiently distinct modal characteristics. In Supplemental Section \ref{supp:spatial_modes}, we show that TDMD in a Fourier transform spectrometer is also capable of separating higher order spatial modes of the same polarization, and of more than 2 modes simultaneously. Derivative-based interferometric modal splitting also applies with respect to stimuli other than temperature, and could also be realized using modal differences in the electro-optic, magnetic-optic, or stress-optic coefficients. In Supplemental Section \ref{supp:electro_optic}, we discuss the broadest generalization of the requirements needed for interferometric modal splitting to be observed, and show that frequency shifting using an electro-optic modulator also exhibits interferometric modal splitting. Though the phenomenon observed is quite different from that used for the dual-polarization spectrometer here, it nonetheless represents another example where modes can be separated by harnessing modal differences rather than engineering to avoid them.

    In conclusion, we have developed the theory behind a dual-polarization Fourier transform spectrometer and experimentally demonstrated it with a diverse set of input light sources. The spectrometer features numerous improvements to previous single-polarization Fourier transform spectrometers in its calibration and operation that enhance performance (Supplementary Sections \ref{supp:setup}-\ref{supp:push_pull}). We generalize the fundamental concept behind the two-polarization capability of the spectrometer, and suggest how this mode splitting capability can be utilized by more advanced spectrometer architectures. All such methods serve as examples of the general idea of interferometric mode separation, which we believe could see widespread use for other techniques, stimuli, and types of orthogonal modes.
\section{Methods}
    \subsection{Integrated photonic chip fabrication and design}
        The spectrometers demonstrated here were fabricated in a standard 220 nm SOI multi-project wafer run at a commercial foundry (Applied Nanotools). This waveguide platform uses a 2 \textmu m thick buried thermal silicon dioxide layer, 220 nm silicon waveguide layer, and a 2.2 \textmu m top PECVD silicon dioxide cladding layer. Heaters are comprised of a 200 nm thick TiW layer, while low-resistance routing uses a 200 nm TiW/500nm Al bilayer. The spectrometers have a footprint of approximately 1750 x 800 \textmu m, and waveguide lengths of 3 cm in each arm. Standard PDK components are used for the edge-couplers and 1x2 50/50 MMI splitters. The TiW heaters used above each spectrometer arm have a width of 25 \textmu m and gaps of 5 \textmu m. 50 \textmu m wide thermal isolation trenches (visible in Fig. 3a) are used to reduce thermal cross-talk between interferometer arms and improve tuning efficiency. When the heaters are driven at the maximum power of 6W, the waveguide temperature is estimated to reach about 90-100 K above room temperature (Supplementary Section \ref{supp:ring}).
    \subsection{Experimental setup and data processing}
        The setup shown in Supplemental Section \ref{supp:setup}, which is a more detailed and thorough version of Fig. 4i, was used for most experiments in this work. The light sources used included two external cavity tunable diode lasers (Santec TSL-550, New Focus Venturi 6600), an amplifed spontaneous emission (ASE) source (Amonics ALS-CL-15-B), and a superluminescent diode (Thorlabs S5FC1005S). A polarization maintaining 1x2 fiber coupler (Thorlabs PMC1550-50B-APC) was used to combine sources. An inline polarizer (AFW Technologies ILP-15-2-SA) was used in some measurements to linearly polarize sources. Cleaved PM-1550 and SMF-28 fiber were used to couple light on and off the chip, respectively. Power-ramping of the interferometer heaters was performed using a 2-channel power supply (Keysight B2962A) and four electrical needle probes. Light collected from the chip was measured using an Agilent 81635A power meter. Reference measurements were performed using two fiber-collimators, a free-space Glan-Thompson polarizer, and an OSA (Anritsu MS9740B), as shown in Fig. 4a. After each measurement using the on-chip spectrometer, fibers were switched to the collimator setup. Two spectra were obtained using the OSA, each with the Glan-Thompson polarizer oriented along a different principal axis of the PM fiber connected to the first fiber-collimator.

        Mode simulations were performed in ANSYS Lumerical MODE (finite difference frequency domain method) and also in Python using the Femwell library (finite element method). MATLAB was used to control all equipment and perform data analysis. See Supplemental Section \ref{supp:data_processing} for details about the data processing steps. 

    \subsection{Data and code availability}
        All data and code used in this paper and supplement have been uploaded to figshare during submission of this manuscript, and will also be made available at \url{https://github.com/nrubinlab/integrated-pol-spectrometer}
\section{Acknowledgments}
We would like to thank the staff of the UC San Diego Makerspace for their assistance in machining mechanical components used in the experimental setups in this work. We would also like to thank Maribel Montero and all of UCSD's nano3 cleanroom staff for assistance in fabricating initial prototypes of the spectrometer demonstrated in this work.
        
\section{Supplementary Information Sections}
\begin{enumerate}
        \item \label{supp:nufft} Single-mode spectrometer analysis and NUDFT extraction method
        \item \label{supp:separability} Two-mode spectrometer analysis and condition for broadband polarization separation
        \item \label{supp:ng_resolution} Derivation of expression for resolution
        \item \label{supp:setup} Experimental setup details
        \item \label{supp:data_processing} Data processing flow
        \item \label{supp:push_pull} Importance of push/pull heater driving
        
        \item \label{supp:widths} Impact of waveguide geometry on spectrometer performance
        \item \label{supp:ring} Ring resonator experiment to estimate temperature rise in spectrometer
        \item \label{supp:er} Additional plots illustrating polarization extinction ratio
        \item \label{supp:other_spectrometers} Interferometric modal splitting in other spectrometer architectures
        \item \label{supp:spatial_modes} Separation of higher-order modes in a Fourier transform spectrometer
        \item \label{supp:electro_optic} Generalization of interferometric modal splitting and electro-optic splitting demonstration
    \end{enumerate}
\printbibliography

\end{document}